\documentclass[aps,prb,twocolumn,groupedaddress,showpacs,floatfix]{revtex4}
\usepackage{amsmath}
\usepackage{graphicx}
\usepackage{bm}

\begin{document}
\title{Comment on ``Comparison of the $Cc$ and $R3c$ space groups 
for the superlattice phase of Pb(Zr$_{0.52}$Ti$_{0.48}$)O$_3$''}

\author{J. Frantti}
\affiliation{Materials and Structures Laboratory, Tokyo Institute of
Technology, 4259 Nagatsuta, Midori-ku, Yokohama, 226-8503, Japan}
\email{jfrantt1@msl.titech.ac.jp}

\date{\today}

\begin{abstract}
The two-phase (space groups $Cc$ and $Cm$) structural model recently 
proposed for the modeling of the neutron powder diffaction pattern 
collected on 
Pb(Zr$_x$Ti$_{1-x}$)O$_3$ (PZT) sample with $x\approx 0.52$ 
[R. Ranjan, A. K. Singh, Ragini, and D. Pandey. \textit{Phys. Rev. B} 
\textbf{71}, 092101 (2005)] is shown to neglect the $hkl$ 
dependent (anisotropic) line broadening. The most serious problem with 
this model is that it assigns octahedral tilts to a wrong phase. 
Instead of correctly taking 
the anisotropic line broadening into account, this model used low 
symmetry phases to minimize the residuals during Rietveld refinement. 
The essential features of a model taking the $hkl$ dependent line 
broadening into account are summarized. It has already been reported 
that once the anisotropic line broadening, revealed by high resolution 
neutron powder diffraction instrument, is correctly taken into account 
the model with $R3c$ and $Cm$ space group symmetries describes all 
Bragg peaks and their intensities well [J. Frantti, S. Eriksson, S. 
Hull, V. Lantto, H. Rundl\"of, and M. Kakihana. \textit{J. Phys.: 
Condens. Matter} \textbf{15}, 6031 (2003).]. It was further shown 
that the $Cm+R3c$ model 
is consistent with the structural features observed at other compositions 
and temperatures, which is particularly important in the vicinity of the 
phase boundary. Also problems related to sample preparation and data 
collection are pointed out. 
\end{abstract}

\maketitle

Recently, based on previously published neutron powder diffraction data 
\cite{Ranjan3,Hatch}, a model which used two monoclinic phases 
(space group symmetries $Cc$ and $Cm$) was proposed to describe 
the structure of 
Pb(Zr$_x$Ti$_{1-x}$)O$_3$ (PZT) with $x\approx 0.52$ \cite{Ranjan}. This 
model assigned a space group symmetries $P4mm$ (at room temperature) or $Cc$ 
(at low temperature) to the phase which was assigned to $Cm$ symmetry in 
refs. \onlinecite{Noheda} and \onlinecite{FranttiPRB}. In addition, the phase, 
which traditionally has been assigned to the rhombohedral symmetry, was 
assigned to $Cm$ symmetry. To understand the problems related to such a model 
and the reason why the use of space group $Cc$ was previously rejected 
\cite{FranttiPRB} it is necessary to summarize the crucial role of the $hkl$ 
dependent line broadening for Rietveld refinement in the case of these 
materials. There is a high risk that erroneous space group symmetry 
assignments result in once this line broadening is compensated for by 
reducing the space group symmetry. For example, one can make the difference 
between measured and calculated diffraction intensities arbitrarily small 
by simply using space group $P1$ and increasing the primitive cell size 
until the difference vanishes below the desired value. In our opinion, a 
structure model which 
reduces the symmetries to 'model' the $hkl$ dependent line broadening takes 
the disorder into account in a wrong way, since in this way the disorder is 
assigned to be a periodical disturbance extending through the crystal, 
obeying the space group symmetry. However, the $hkl$-dependent line broadening 
is \emph{an inherent} property of PZT powders, and is already seen in 
tetragonal Ti rich compositions. The same behaviour was observed in a closely 
related Pb(Hf$_x$Ti$_{1-x}$)O$_3$ (PHT) system with $0.10\leq x \leq 0.40$, 
where the peak widths of the $00l$ reflections were \emph{twice} as large as 
the widths of $h00$ reflections\cite{FranttiSub}. It must be emphasized that 
no signs of symmetry lowering from the $P4mm$ symmetry was observed in this 
high- resolution neutron powder diffraction study. Light scattering 
experiments revealed that both in the case of PZT and PHT samples 
deviations from the average symmetry were observed 
\cite{FranttiPRB2,FranttiJJAP2,FranttiBoston}. Although one could decrease 
the average symmetry to take these observations into account, a more 
realistic model is to assume that the \emph{average symmetry} over a 
length scale of a few hundred nanometers is $P4mm$, and deviations from 
this symmetry occur in 
a local scale (of a few unit cells). The essential point here, 
too, is that these deviations are not periodical! It is also important 
to note that the $Cm$ phase was identified by studying the peak split 
of the pseudo-cubic $110$ reflections \cite{Noheda}. This peak split 
was beyond the resolution of the instrument used in ref. 
\onlinecite{Ranjan} 
and there is no way to recover the lost information. It does not matter 
how many 'new' refinements are carried out for the low-resolution data 
collected with insufficient counting time (note that neutron powder 
diffraction data in refs. \onlinecite{Ranjan3,Hatch} and 
\onlinecite{Ranjan} are the same). 
To solve the problem related to average symmetries it is necessary to carry 
out experiments with an appropriate instrument and carefully prepared sample.
The role of the spatial composition 
variation (which in practice cannot be eliminated) in the vicinity of 
the phase boundary is an old and still a valid explanation for the two 
phase 'co-existences' and also partially explains the $hkl$-dependent 
line broadening. It was also worth of trying to find a simple model which 
is consistent with other compositions and 
temperatures\cite{FranttiPRB,FranttiJPCM}. These aspects 
are shortly reviewed below.

\begin{figure}
\includegraphics[width=8.6cm]{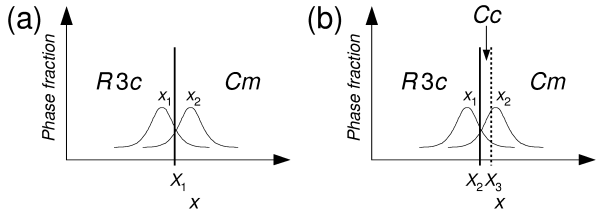}
\caption{\label{CompVar} The consequences of the spatial composition 
variation (at fixed temperature and pressure) in the case of two samples 
with average compositions $x_1$ and $x_2$ in the vicinity of (a) the phase 
boundary separating $R3c$ and $Cm$ phases (at $X_1$) and (b) two phase 
boundaries separating $R3c$ and $Cc$ (at $X_2$) and $Cc$ and $Cm$ phases 
(at $X_3$).}
\end{figure}

\paragraph{Spatial composition variation.} Traditionally, the 
two-phase 'co-existence', observed in room temperature PZT in the 
vicinity of the morphotropic phase boundary (MPB), has been 
explained by the composition variation. In the context of low 
temperature symmetries the role of spatial composition variation was 
discussed in refs. \onlinecite{FranttiPRB} and \onlinecite{FranttiJPCM}. 
The crystal symmetry of Zr rich PZT ceramics is $R3c$ 
\cite{Michel,Corker,FranttiJPCM}, except for the compositions with 
$x \approx 1$. 
Since spatial composition variation cannot be completely eliminated, there 
must exist two phases in the vicinity of MPB. To allow the existence of $Cc$ 
phase necessitates that there should be a narrow region in the $x-T$ plane 
were this phase is stable or metastable.
Now, if we were to explain the existence of $Cc$ phase, two phase boundaries 
located somewhere between $0.52 \leq x \leq 0.54$ should be assumed to exists, 
see Fig. \ref{CompVar} (b). 
This in turn, once the spatial composition variation is taken into account, 
leads to three phase 'co-existence'. We preferred the simplest $Cm+R3c$ 
model (corresponding to Fig. \ref{CompVar} (a)), since it was able to 
explain the experimental observations in simplest terms. At low temperature 
(4 K\cite{FranttiJPCM} or 10 K \cite{FranttiPRB}) the phase fraction of 
the $Cm$ phase was monotonically decreasing with $x$ increasing from $0.52$ 
to $0.54$, which in turn implies that two-phase 'co-existence' is 
predominantly due to the spatial composition variation. Other factors 
include stresses, which probably exist in the grain boundaries. 
It is worth to point out that the model proposed in ref. 
\onlinecite{Ranjan} was 
ruled out also in a recent paper, see the last paragraph 
in ref. \onlinecite{Cox}. 
After all, the only difference between our model and the model proposed in 
ref. \onlinecite{Cox} is that we used $R3c$ symmetry (corresponding to two 
lattice parameters and four atomic coordinates), while they used $Cc$ 
symmetry (corresponding to four lattice parameters and twelve atomic 
coordinates, if constraints are not used), a subgroup of $R3c$ space group. 
Thus, the model proposed by us and the model proposed in ref. 
\onlinecite{Cox} assigned octahedral tilts (evidenced by superlattice 
reflections) to the same phase. Octahedral tilts are crucial once the 
structural changes versus temperature, composition or pressure are studied. 
Somewhat surprisingly, the space 
group symmetries of the two phases were rather arbitrarily interchanged in 
refs. \onlinecite{Ranjan} and \onlinecite{Ranjan2}. This model 
immediately runs into trouble, as the octahedral tilts are assigned to a 
wrong phase, see below. 

In addition, there 
have been reports claiming that dielectric measurements support $Cc$ 
phase (for example, see ref. \onlinecite{Ranjan3}). In this context we note 
that the samples studied in 
refs. \onlinecite{Ranjan} and \onlinecite{Ranjan2} contained a nonidentified 
impurity phase(s), as was revealed by the peaks at around 28 and 35 two-theta 
degrees (corresponding to $d$ spacings 3.18 \AA\mbox{  }and 2.56 \AA, 
respectively). This implies that the composition was not well known and 
there is no way to avoid compositional and structural inhomogeneities. 
Although this non-perovskite phase was clearly observable, it was neglected 
in the Rietveld refinement model considered in ref. \onlinecite{Ranjan2}, 
which in turn results in an error in the structural parameters of the 
perovskite 
phase(s). This is related to the fact that the 
diffraction pattern shown in ref. \onlinecite{Ranjan2} (reported to 
have $x=0.52$) is reminiscent to our diffraction pattern with $x=0.53$. 
In contrast, our diffraction pattern with $x=0.52$ (see ref. 
\onlinecite{FranttiJJAP}) and the diffraction pattern of the $x=0.52$ 
sample reported in refs. \onlinecite{Cox} and \onlinecite{Noheda} were 
reminiscent. It should also be understood that dielectric measurements  
do not suit for a space group determination. Although the dielectric 
measurements can provide valuable information in the case of single 
phase samples, particular care is necessary once multiphase samples are 
studied. Anomalies observed in a dielectric constant versus temperature 
measured from a \emph{two or three phase} ceramic bulk samples 
(see ref. \onlinecite{Ragini3}) do 
not provide reliable evidence that one of the phases has 
undergone a phase transition from $Cm$ to $Cc$ phase. 

\paragraph{Anisotropic line broadening.} The anisotropic line broadening 
was ascribed to the spatial 
composition variation resulting in 'microstrain' (\emph{i.e.}, Zr 
substitution for Ti creates local strains which in turn contributes 
to the $hkl$ dependent line broadening)\cite{FranttiJPCM}.  In 
addition, rather strong diffuse scattering is commonly 
observed between the $h00$ and $00h$ Bragg reflections, which was assigned to 
Pb ions displaced toward $\langle 110 \rangle$ directions. This latter feature 
is somewhat puzzling, as it adds intensity to certain diffraction peaks, which 
results in \emph{an asymmetric profile}. For example, the larger and smaller 
$d$-spacings sides of the pseudo-cubic $200$ and $002$\mbox{  }reflections, 
respectively, gain intensity. In ref. \onlinecite{Ranjan} $Cm$ phase was 
practically used to model the intensity not only due to the crystalline phase, 
but also due to the diffuse scattering and due to the main crystalline phase. 
Despite the increased number of refined parameters, the residuals were still 
rather high and the differences between different models compared were 
marginal (for instance, $R_{exp}$ was lowest for the $R3c+Cm$ model which 
was rejected in ref. \onlinecite{Ranjan}).

\begin{figure}
\includegraphics[width=8.6cm]{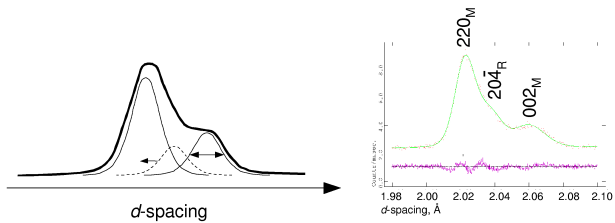}
\caption{\label{Asymmetry} Left panel: Schematic illustration of the 
way how the neglect of anisotropic line broadening affects the 
position of the rhombohedral peaks: once the rightmost peak broadens, 
it pushes the rhombohedral peak (middle) towards smaller $d$-spacing. 
Righ panel: Observed and computed high resolution neutron powder 
diffraction profile collected on PZT with $x=0.53$ at 4 K. Pseudo-cubic 
$200$\mbox{  }reflection region is shown. Note that the $002_M$ peak 
($Cm$ phase) at around 2.06 \AA\mbox{  }is significantly broader than 
the $220_M$ ($Cm$ phase) peak at around 2.02 \AA. Now, the position 
of the rhombohedral $20\bar{4}_R$ peak depends on the way the line 
broadening is taken into account.}
\end{figure}

Although a reasonable structure refinement for medium resolution data 
collected on PZT samples with $x\leq 0.50$ could be obtained using a 
lineshape which ignores anisotropic line broadening, the situation was 
quite different for high resolution facilities, particularly once the 
data was collected on two phase samples (the present case). In such 
a case it was essential to use an appropriate profile function. 
\emph{Now, if the anisotropic line broadening is neglected}, the 
fit in the case of certain weak peaks becomes slightly worse (as was 
observed to be the case of weak superlattice reflections, which are 
less weighted in the refinements). The reason for this is illustrated 
in Fig. \ref{Asymmetry}: 
in order to improve the fit corresponding to the strong pseudo-cubic 
$200$ reflections, during the refinement the position of $R3c$ 
reflection is shifted toward higher $d$-spacings, which in turn 
resulted in a small shift of the weak peaks at 2.44 and 1.06 
\AA\mbox{  }(in a case of the data shown in ref. \onlinecite{Ranjan} 
the latter peak was almost at the level of noise and the $Cc+Cm$ model 
assigns more reflections than there are data points in this region). 
The anisotropic 
line broadening was not limited to the $d$-spacing at around 2 \AA, 
and similar mechanism was seen at other $d$-spacings. 

\paragraph{Previous high resolution neutron powder diffraction results.}
The anisotropic line broadening was the reason to carry out a 
subsequent study using a high resolution neutron powder diffraction 
instrument (which reveals the anisotropic line broadening particularly 
clearly and 
allows a distinction between average symmetry lowering and a defect 
related anisotropic line broadening to be done, in contrast to the 
low- or medium resolution constant wavelength neutron diffraction 
facilities)\cite{FranttiJPCM}. In addition, we increased Zr content, 
$x=0.54$\cite{FranttiJPCM} (although it definitely was possible to 
fit \emph{all} reflections using the $Cm+R3c$ model also in the 
case of $x=0.52$ and $x=0.53$\mbox{  }samples). This allowed a more 
reliable refinement and symmetry identification to be done by studying 
the \emph{changes in phase fractions versus temperature}. Importantly, 
for these compositions it has been found that oxygen octahedra tilts 
increase with increasing $x$ (ref. \onlinecite{Corker}) and decreasing 
temperature (this feature is discussed in refs. \onlinecite{Thomas} 
and \onlinecite{FranttiPRB}). We also note that previously $Cc$ 
was proposed to be space group symmetry corresponding to high 
isotropic pressure\cite{Haines}, whereas 
our work was concentrating on the determination of space group 
symmetries 
versus composition and temperature at ambient pressure. Now, PZT 
sample with $x=0.54$ provided a test for 
clarifying which phase is the preferred one at low temperature. To 
model the peak profiles, GSAS\cite{GSAS} lineshape 4 by Stephens 
\cite{Stephens} was used. As was seen in ref. \onlinecite{FranttiJPCM}, 
once the crystal contracts with decreasing temperature the $R3c$ phase 
was favoured (and $Cm$ phase did not transform to $Cc$ phase, as one 
could expect from the model proposed in ref. \onlinecite{Ranjan}). 
Thus, when the oxygen octahedra had almost no chance to contract, 
they were tilt towards opposite directions along the pseudo-cubic 
$111$ axis (tilt system $a^-a^-a^-$). This tilt results in a decrease 
of the volume of cuboctahedra around Pb ions. Changes evidencing  
$Cc$ symmetry (and the corresponding tilt system) were not observed. 
\emph{Indeed, the phase fraction of $Cm$ phase (which was assigned 
to $Cc$ symmetry in ref. \onlinecite{Ranjan}) decreased with 
decreasing temperature, whereas $R3c$ phase fraction significantly 
increased with decreasing temperature. This can be confirmed even by a 
naked eye, although the results of refinements were also given. The 
intensity of the peak at around 1.06 \AA\mbox{  }was increasing with 
decreasing temperature, which further confirmed that its origin is 
the phase assigned to $R3c$ phase. Also the peak at around 
1.06 \AA\mbox{  }was well fit by $R3c$ symmetry}. 

As a summary, we do not find support for $Cm+Cc$ model proposed in ref. 
\onlinecite{Ranjan}. Instead, we still consider that $Cm+R3c$ provides 
a sufficient low temperature structural model for PZT with composition 
in the vicinity of the morphotropic  phase boundary.




\begin{thebibliography}{99}
\bibitem{Ranjan3}  R. Ranjan, S. K. Mishra, D. Pandey and K. Kennedy. 
\textit{Phys. Rev. B} \textbf{65}, 060102 (2002).
\bibitem{Hatch}  D. M. Hatch, H. T. Stokes, R. Ranjan, Ragini, 
S. K. Mishra, D. Pandey, and B. J. Kennedy, 
\textit{Phys. Rev. B} \textbf{65}, 212101 (2002).
\bibitem{Ranjan} R. Ranjan, A. K. Singh, Ragini, and D. Pandey. 
\textit{Phys. Rev. B} \textbf{71}, 092101 (2005).
\bibitem{Noheda} B. Noheda, D. E. Cox, G. Shirane, J. A. Gonzalo, 
L. E. Cross and S-E. Park. \textit{Appl. Phys. Lett.} \textbf{74}, 
2059 (1999); B. Noheda, J. A. Gonzalo, L. E. Cross, R. Guo, S-E. Park, 
D. E. Cox, and G. Shirane. \textit{Phys. Rev. B} \textbf{61}, 8687 (2000).
\bibitem{FranttiPRB} J. Frantti, S. Ivanov, S. Eriksson, H. Rundl\"of, 
V. Lantto, J. Lappalainen, and M. Kakihana. \textit{Phys. Rev. B} 
\textbf{66}, 064108 (2002).
\bibitem{FranttiSub} J. Frantti, Y. Fujioka, S. Eriksson, S. Hull and 
M. Kakihana. Neutron powder diffraction study of Pb(Hf$_x$Ti$_{1-x}$)O$_3$ 
ceramics ($0.10 \leq x \leq 0.50$), submitted. 
\bibitem{FranttiPRB2} J. Frantti, V. Lantto, S. Nishio and M. Kakihana.
\textit{Phys. Rev. B} \textbf{59}, 12 (1999).
\bibitem {FranttiJJAP2} J. Frantti, J. Lappalainen, S. Eriksson, 
V. Lantto, S. Nishio, M. Kakihana, S. Ivanov, and H. Rundl\"of. 
\textit{Jpn. J. Appl. Phys.} \textbf{38}, 5679 (1999).
\bibitem{FranttiBoston} J. Frantti, Y. Fujioka, S. Eriksson, V. Lantto, 
 and M. Kakihana. \textit{Journal of Electroceramics} \textbf{13}, 299 (2004). 
\bibitem{Ranjan2} Ragini, R. Ranjan, S. K. Mishra, and D. Pandey, 
\textit{J. Appl. Phys.} \textbf{92}, 3266 (2002).
\bibitem {FranttiJJAP} J. Frantti, J. Lappalainen, S. Eriksson, 
V. Lantto, S. Nishio, M. Kakihana, S. Ivanov, and H. Rundl\"of. 
\textit{Jpn. J. Appl. Phys.} \textbf{39}, 5697 (2000).
\bibitem{FranttiJPCM} J. Frantti, S. Eriksson, S. Hull, V. Lantto, 
H. Rundl\"of, and M. Kakihana. \textit{J. Phys.: Condens. Matter} 
\textbf{15}, 6031 (2003).
\bibitem{Michel} C. Michel, J. M. Moreau, G. D. Achenbach, R. Gerson, and 
W. J. James, \textit{Solid State Commun.} \textbf{7}, 865 (1969).
\bibitem{Corker} D. L. Corker, A. M. Glazer, R. W. Whatmore, A. Stallard, 
and F. Fauth. \textit{J. Phys. Condens. Matter} \textbf{10}, 6251 (1998).
\bibitem{Cox} D. E. Cox, B. Noheda, and G. Shirane. \textit{Phys. Rev. B} 
\textbf{71}, 134110 (2005).
\bibitem{Haines} J. Rouquette, J. Haines, V. Bornand, M. Pintard, Ph. 
Papet, W. G. Marshall, and S. Hull. \textit{Phys. Rev. B} \textbf{71}, 
024112 (2005).
\bibitem{Ragini3} Ragini, S. K. Mishra, D. Pandey, H. Lemmens, and G. 
Van Tendeloo. \textit{Phys. Rev. B} \textbf{64}, 
054101 (2001).
\bibitem{Thomas} N. W. Thomas and A. Beitollahi. 
\textit{Acta Crystallorg., Sect. B: Struct. Sci.} \textbf{50}, 549 (1994).
\bibitem{GSAS} A. C. Larson and R. B. Von Dreele \textit{General 
Structure Analysis System}, LANSCE MS-H805, Los Alamos National 
Laboratory, Los Alamos, NM 87545 (2000).
\bibitem{Stephens} P. W. Stephens.  J. Appl. Crystallogr. \textbf{32} 
281 (1999).
\end{thebibliography}
\end{document}